\documentclass[aps,prb,twocolumn,10pt,longbibliography,superscriptaddress]{revtex4-1}

\usepackage{amsmath}
\usepackage{amsfonts}
\usepackage{amssymb}{\tiny}
\usepackage{mathtools}
\usepackage{times}
\usepackage{hyperref}
\usepackage{bm}
\usepackage{cases}
\usepackage{graphicx}
\usepackage{xcolor}

\begin{document}

\preprint{APS/123-QED}

\title{Chiral spin liquids with crystalline $\mathbb{Z}_2$ gauge order in a three-dimensional Kitaev model}

\author{Petr A. Mishchenko}
\affiliation{Department of Applied Physics, University of Tokyo, Tokyo 113-8656, Japan}
\author{Yasuyuki Kato}
\affiliation{Department of Applied Physics, University of Tokyo, Tokyo 113-8656, Japan}
\author{Kevin O'Brien}
\affiliation{Institute for Theoretical Physics, University of Cologne, 50937 Cologne, Germany}
\author{Troels A. Bojesen}
\affiliation{Department of Applied Physics, University of Tokyo, Tokyo 113-8656, Japan}
\author{Tim Eschmann}
\affiliation{Institute for Theoretical Physics, University of Cologne, 50937 Cologne, Germany}
\author{Maria Hermanns}
\affiliation{Department of Physics, Stockholm University, SE 106 91 Stockholm, Sweden}
\affiliation{Nordita, KTH Royal Institute of Technology and Stockholm University, SE 106 91 Stockholm, Sweden}
\author{Simon Trebst}
\affiliation{Institute for Theoretical Physics, University of Cologne, 50937 Cologne, Germany}
\author{Yukitoshi Motome}
\affiliation{Department of Applied Physics, University of Tokyo, Tokyo 113-8656, Japan}

\date{\today}

\begin{abstract}
Chiral spin liquids (CSLs) are  time-reversal symmetry breaking ground states of frustrated quantum magnets that show no long-range
magnetic ordering, but instead exhibit topological order and fractional excitations. Their realization in simple and tractable microscopic
models has, however, remained an open challenge for almost two decades until it was realized that Kitaev models on lattices with odd-length
loops are natural hosts for such states, even in the absence of a time-reversal symmetry breaking magnetic field. Here we report on the
formation of CSLs in a three-dimensional Kitaev model, which differ from their widely studied two-dimensional counterparts, namely, they
exhibit a crystalline ordering of the $\mathbb{Z}_2$ gauge fluxes and thereby break some of the underlying lattice symmetries. We study the
formation of these unconventional CSLs via extensive quantum Monte Carlo simulations and demonstrate that they are separated from the
featureless paramagnet at high temperatures by a single first-order transition at which both time-reversal and lattice symmetries are
simultaneously broken. Using variational approaches for the ground state, we explore the effect of varying the Kitaev couplings and find at
least five distinct CSL phases, all of which possess crystalline ordering of the $\mathbb{Z}_2$ gauge fluxes. For some of these phases, the
complementary itinerant Majorana fermions exhibit gapless band structures with topological features such as Weyl nodes or nodal lines in
the bulk and Fermi arc or drumhead surface states.
\end{abstract}

\maketitle

%
\section{Introduction}
%

Among quantum spin liquids  -- entangled states of matter that defy conventional magnetic order~\cite{balents_nature_464_2010,
savary_rpp_80_2017} -- chiral spin liquids (CSLs) form a particularly intriguing subset. Originally conceptualized as bosonic analogues of
fractional quantum Hall states~\cite{kalmeyer_prl_59_1987}, they have long been sought-after as ground states of frustrated quantum
magnets~\cite{lacroix}. While initial proposals have concentrated on resonating valence bond (RVB)
states~\cite{anderson_mater_res_bull_8_1973,fazekas_phil_mag_30_1974}, suspected to form as ground states of geometrically frustrated
Heisenberg antiferromagnets and relevant to high-temperature ($T$) superconductivity~\cite{anderson_science_235_1987,wen_prb_39_1989},
it was only recently recognized that they could be firmly established as ground states of microscopic model Hamiltonians. On the numerical
side, decisive progress has been made via the calculation of modular matrices~\cite{Rowell2009,Bruillard2016} from minimally entangled
states~\cite{Zhang2012} in density matrix renormalization group (DMRG) calculations~\cite{Stoudenmire2012} that allowed to identify CSL
ground states in a number of kagome antiferromagnets with interactions beyond the nearest neighbor exchange~\cite{messo_prl_108_2012,
bauer_ncom_5_2014,gong_srep_4_2014}. On the analytical side, Kitaev pointed out that extensions of his spin model to lattices with
elementary loops of odd length can host CSL ground states~\cite{kitaev_ann_phys_321_2006}, as first demonstrated for a two-dimensional (2D)
triangle-honeycomb lattice by Yao and Kivelson~\cite{yao_prl_99_2007}. This can be rationalized by considering the hopping of the itinerant
Majorana fermions in Kitaev's exact analytical solution~\cite{kitaev_ann_phys_321_2006}, relevant to any tricoordinated lattice geometry --
every hop between two sites is accompanied by a complex amplitude ${\mathrm i}$, which for an odd-length loop results in a (positive or
negative) multiple of ${\mathrm i}$ or, equivalently, a flux of $\pm \pi/2$ per plaquette. In the ground state, this flux will be fixed to
one of the two values implying that time-reversal symmetry, which connects the two possible choices, will be broken. As such the formation
of a CSL at low $T$ must be separated from the featureless paramagnet at high $T$ through a finite-$T$ phase transition at which
time-reversal symmetry is spontaneously broken. That this is indeed the case has been demonstrated via quantum Monte Carlo (QMC)
simulations~\cite{nasu_prl_113_2014} for various 2D lattice geometries~\cite{nasu_prl_115_2015,dwivedi_prb_98_2018}, which have also
confirmed the universality class of this continuous transition to be of Ising-type.

In this manuscript, we consider three-dimensional (3D) CSLs which in contrast to their 2D counterparts have remained considerably less
studied. Turning to 3D variants of the Kitaev model~\cite{obrien_prb_93_2016}, we are particularly interested in their thermodynamic
behavior -- the motivation for this comes from the observation that the underlying $\mathbb{Z}_2$ gauge structure of the Kitaev model
exhibits fundamentally different behavior in two and three spatial dimensions. While in two dimensions the elementary $\mathbb{Z}_2$ flux
excitations, also referred to as visons, are point-like objects, they are closed loops in three spatial dimensions. This difference points
to the existence of a topological phase transition, at which the loops proliferate and ``break open'' into line-like objects spanning the
entire system length~\cite{Senthil2000}. QMC simulations for the 3D Kitaev models on the so-called hyperhoneycomb~\cite{nasu_prl_113_2014}
and hyperoctagon lattices~\cite{mishchenko_prb_96_2017}, lattices with elementary plaquettes of even length 10, have indeed observed such a
finite-$T$ transition, whose continuous nature is consistent with the expected inverted-Ising universality class. This raises the question
whether, for a 3D Kitaev model on a lattice geometry with odd-length plaquettes, there are multiple phase transitions associated with the
ordering of the $\mathbb{Z}_2$ gauge structure and the spontaneous breaking of time-reversal symmetry, respectively.

To address this question, we consider the 3D Kitaev model on the so-called hypernonagon lattice, a 3D lattice geometry with elementary
plaquettes of length 9, as illustrated in Fig.~\ref{fig:fig_1}. Guided by a previous study~\cite{kato_prb_96_2017} concentrating on the
limit of strong anisotropy in the exchange couplings, in which perturbation theory can be applied to derive an effective model, we further
take into account the possibility that the low-$T$ $\mathbb{Z}_2$ gauge configuration exhibits a nonuniform flux pattern by considering
exchange couplings beyond the anisotropic limits. This gives an even larger potential for thermodynamic phase transitions at which
{\em multiple} symmetries are broken spontaneously and as such evade a conventional description in terms of the Landau-Ginzburg-Wilson
paradigm. To explore this physics we run extensive QMC simulations of the hypernonagon Kitaev model -- an exacting endeavor due to the
relatively large sizes resulting from a unit cell with 12 sites for the hypernonagon lattice that can be pursued only because of a plethora
of algorithmic tricks, including the implementation of the kernel polynomial method~\cite{weisse_prl_102_2009,weisse_rev_mod_phys_78_2006,
mishchenko_prb_96_2017} and optimized parallel tempering schemes~\cite{katzgraber_jsm_03018_2006}. Our numerical results show that there is
only a single finite-$T$ phase transition at which both time-reversal and lattice symmetries are simultaneously broken. This transition
turns out to be of first order where the $\mathbb{Z}_2$ gauge configuration indeed forms a nonuniform flux pattern. In variational
calculations for varying relative strengths of the Kitaev couplings, we find that such nonuniform flux patterns are present in the entire
zero-$T$ phase diagram. In total, these ordering patterns of the $\mathbb{Z}_2$ gauge fluxes allow us to distinguish five distinct phases.
Interestingly, in all but one out of these phases, the crystalline order of the $\mathbb{Z}_2$ gauge fluxes is not commensurate with the
original lattice symmetries, i.e., at least one point-group symmetry of the lattice is broken in addition to time-reversal symmetry.

To round off our analysis of the hypernonagon Kitaev model, we compute the band structure of the complementary fractional excitations --
the itinerant Majorana fermions -- in all the CSL phases. We identify the boundaries between gapped and gapless phases, which do not
necessarily coincide with the phase boundaries between the different flux phases. Finally, we investigate the topological nature of the
Majorana band structures, which is particularly rewarding in the gapless phases for which we find topological features such as Weyl nodes
and nodal lines. These bulk features are accompanied by topologically protected surface states, Fermi arcs and drumhead states. We
characterize these features for all CSL phases at hand.

Our discussion in the remainder of the paper is structured as follows. In Sec.~\ref{sec:model}, we discuss the elementary features of the
hypernonagon lattice and how to uniquely define a Kitaev model for this lattice geometry. Before reporting on our results, we provide a
brief overview of the relevant technical details of our finite-$T$ QMC simulations, zero-$T$ variational calculations, and the
determination of the Majorana band structures in Sec.~\ref{sec:methods}. Results from these calculations are presented in
Secs.~\ref{sec:finite-t} and \ref{sec:zero-t}, concentrating on the thermodynamics and zero-$T$ physics, respectively. A summary and some
concluding remarks are given in Sec.~\ref{sec:summary_remarks}.

%
\section{Model}
\label{sec:model}
%

To set the stage for our discussion, we first introduce the hypernonagon lattice and review its fundamental symmetries. We further discuss
an alternative, but less symmetric basis that allows to capture some of the geometrical aspects of the lattice in more accessible terms.
Finally, we explain how to uniquely define a Kitaev model for this lattice and review the structure of conserved $\mathbb{Z}_2$ fluxes.

%
\subsection{The hypernonagon lattice}
\label{sec:lattice}
%

\begin{figure}[tb]
	\includegraphics[width=\columnwidth]{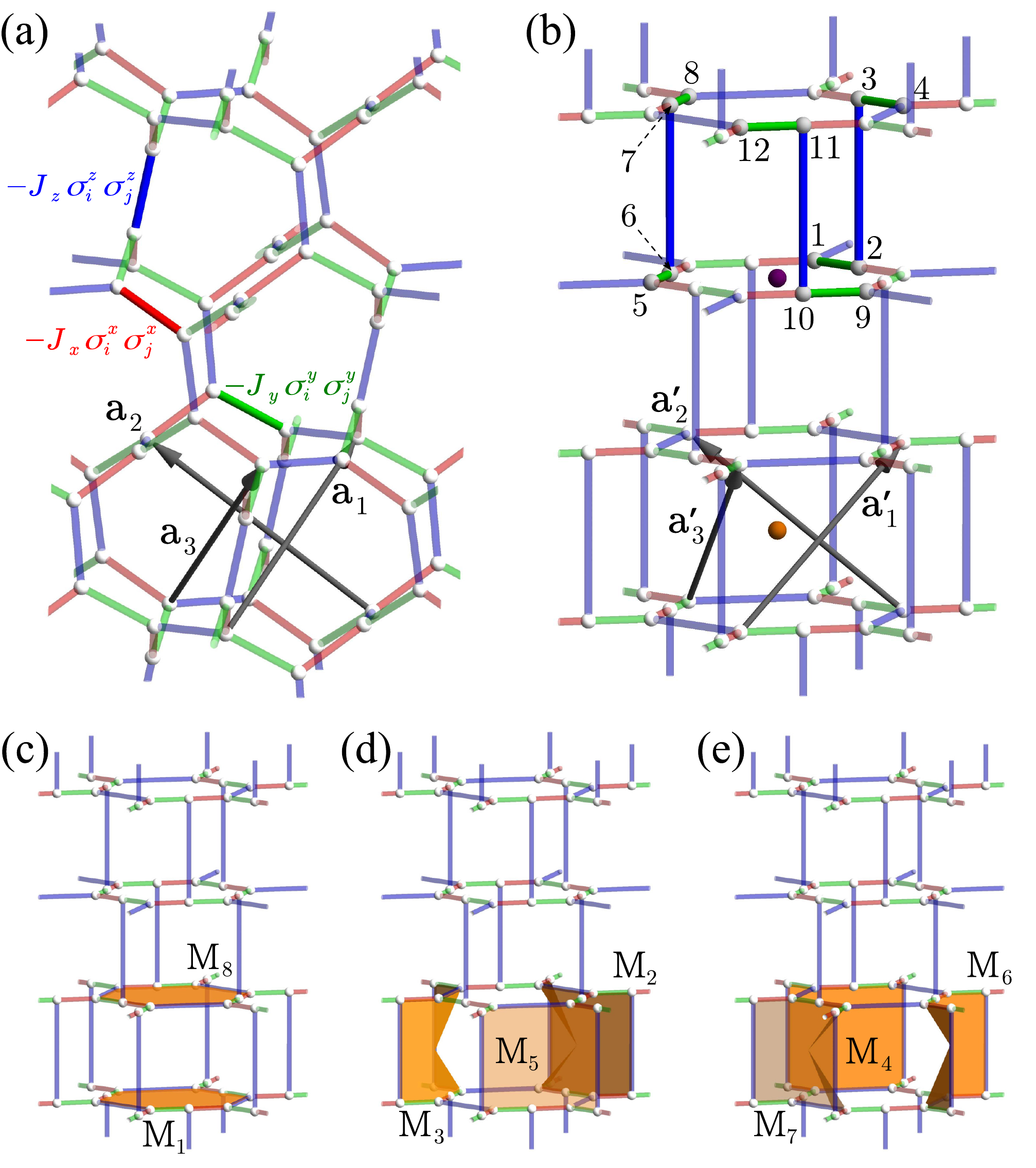}
	\caption{
		Schematic pictures of the Kitaev model on the hypernonagon lattice. (a) The original hypernonagon lattice~\cite{wells} and (b) the
		distorted one for better visibility~\cite{obrien_prb_93_2016}. ${\bf a}_1$, ${\bf a}_2$, and ${\bf a}_3$ (${\bf a}'_1$, ${\bf a}'_2$,
		and ${\bf a}'_3$) are the primitive translation vectors for the original (deformed) lattice. The white spheres represent the lattice
		sites on which the spin-$1/2$ degrees of freedom are defined. The red, green, and blue bonds denote the $x$, $y$, and $z$ bonds of the
		Kitaev model in Eq.~\eqref{eq:kitaev_hamiltonian}, respectively. A unit cell contains 12 sites, as indicated by the numbers 1--12 in
		(b). Two types of inversion centers are denoted by orange and purple spheres in (b). (c)-(e) The eight types of plaquettes M$_p$
		($p = $ 1--8) in a unit cell on which the $\mathbb{Z}_2$ fluxes $W_p$ in Eq.~\eqref{eq:plaquette} are defined. Each plaquette consists
		of 9 sites.
	}
	\label{fig:fig_1}
\end{figure}

The hypernonagon lattice is a 3D lattice structure that retains the tricoordination of every lattice site that is familiar from the 2D
honeycomb lattice. What sets it apart from other tricoordinated lattice structures in three spatial dimensions, which have been extensively
classified by Wells in 1970s~\cite{wells} and reintroduced in the context of 3D Kitaev models recently~\cite{obrien_prb_93_2016}, is that
it exhibits elementary plaquettes with an {\em odd} number of bonds. The fact that these are 9 bonds per plaquettes motivates the
Schl\"afli symbol ``(9,3)a'' in the classification of Wells and the more colloquial name ``hypernonagon'' employed here.

\begin{figure}[tb]
	\includegraphics[width=0.8\columnwidth]{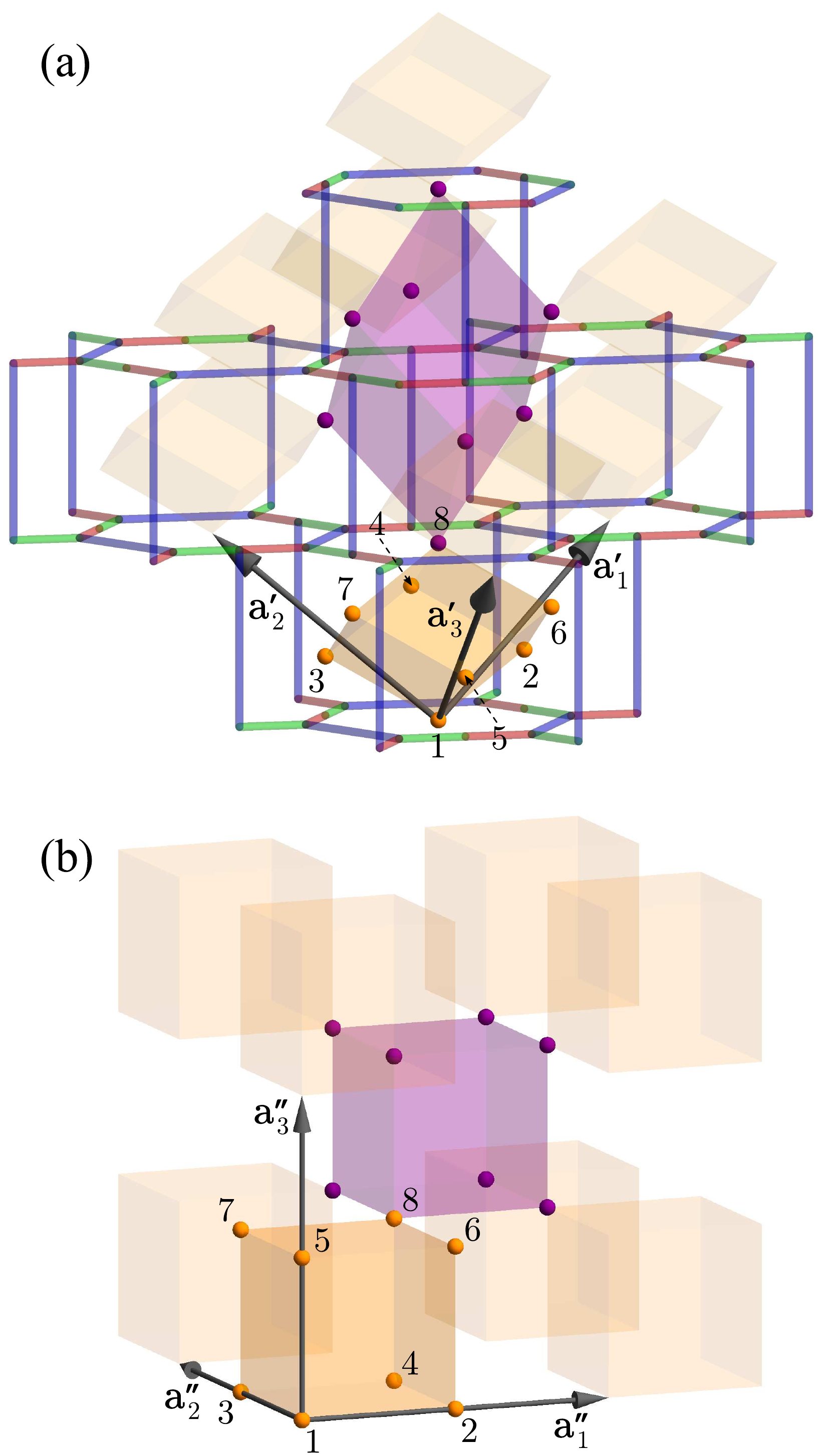}
	\caption{
		3D network of the $\mathbb{Z}_2$ fluxes. (a) The hypernonagon lattice contains two types of closed volumes indicated by the orange and
		purple hexahedra. Both are constructed from the centers of the eight plaquettes, indicated by spheres. The numbers 1--8 indicate the
		plaquettes M$_1$--M$_8$ in Figs.~\ref{fig:fig_1}(c)-\ref{fig:fig_1}(e) on which the $\mathbb{Z}_2$ fluxes $W_p$ are defined. The
		spheres form a distorted cubic lattice. (b) The simple cubic lattice deformed from (a). ${\bf a}'_1$, ${\bf a}'_2$, and ${\bf a}'_3$ in
		(a) [${\bf a}''_1$, ${\bf a}''_2$, and ${\bf a}''_3$ in (b)] are the primitive translation vectors for the deformed (simple) cubic
		lattice.
	}
	\label{fig:fig_2}
\end{figure}

The principal geometrical structure of the hypernonagon lattice is illustrated in Fig.~\ref{fig:fig_1}(a), with a slightly more
accessible, topologically equivalent, but deformed structure shown in Fig.~\ref{fig:fig_1}(b). For this latter version, it is easy to see
that the lattice possesses a $C_3$ symmetry with a rotation axis in the middle of a 12-site loop pointing in the $\hat{e}_z$-direction
(which is parallel to the ${\bf a}_1 + {\bf a}_2 + {\bf a}_3$ direction or the ${\bf a}'_1 + {\bf a}'_2 + {\bf a}'_3$ direction, see
Fig.~\ref{fig:fig_2}). There are three mirror planes spanned by $\hat{e}_z$ and one of primitive translation vectors which cut through the
$z$ bonds (colored in blue in Fig.~\ref{fig:fig_1}) and act to map the $x$ and $y$ bonds (red and green) onto one another by mapping
${\bf a}_2 \leftrightarrow {\bf a}_3$, ${\bf a}_3 \leftrightarrow {\bf a}_1$ or ${\bf a}_1 \leftrightarrow {\bf a}_2$ (the same holds for
${\bf a}'_1$, ${\bf a}'_2$, and ${\bf a}'_3$). Finally, the lattice is inversion symmetric with multiple, unique inversion centers. One
inversion center lies at the centers of the 12-site loop and shown in Fig.~\ref{fig:fig_1}(b) as a purple sphere, whereas the other is
located at the midpoint of the lines connecting the aforementioned inversion centers for the neighboring unit cells and shown in
Fig.~\ref{fig:fig_1}(b) as an orange sphere.

In fact, one can further simplify the lattice geometry when focusing on the elementary plaquettes. As shown in
Figs.~\ref{fig:fig_1}(c)-\ref{fig:fig_1}(e), the lattice structure is composed of eight types of 9-site plaquettes in each unit cell,
M$_1$--M$_8$. They form a closed volume as illustrated by the orange hexahedron in Fig.~\ref{fig:fig_2}(a), where the numbered spheres
represent the eight plaquettes in Figs.~\ref{fig:fig_1}(c)-\ref{fig:fig_1}(e). In addition to the orange hexahedron, one can find a second
closed volume which spans across neighboring unit cells, illustrated by the purple hexahedron in Fig.~\ref{fig:fig_2}(a). Note that the
purple and orange inversion centers in Fig.~\ref{fig:fig_1}(b) correspond to the centers of the purple and orange hexahedra, respectively.
Importantly, these two types of hexahedra form a distorted cubic lattice, which can be further deformed to a simple cubic lattice as shown
in Fig.~\ref{fig:fig_2}(b)~\cite{kato_prb_96_2017,kato_physb_536_2018}.

%
\subsection{Kitaev model and conserved $\mathbb{Z}_2$ fluxes}
\label{sec:flux}
%

In the following, we  will consider a standard Kitaev model
\begin{equation}
\mathcal{H} = -J_x\sum_{\langle i,j \rangle_x} \sigma_i^x \sigma_j^x
              -J_y\sum_{\langle i,j \rangle_y} \sigma_i^y \sigma_j^y
              -J_z\sum_{\langle i,j \rangle_z} \sigma_i^z \sigma_j^z \,,
\label{eq:kitaev_hamiltonian}
\end{equation}
where $\sigma_i^x$, $\sigma_i^y$, and $\sigma_i^z$ are the Pauli matrices describing a spin-$1/2$ degree of freedom at site $i$. The sum
$\langle i,j \rangle_\gamma$ is taken over all the $\gamma$ bonds ($\gamma = x, y, z$, corresponding to three different types of bonds),
and $J_\gamma$ is the exchange constant for the $\gamma$ bonds. Such a Kitaev model can be defined for any tricoordinated lattice, even in
three spatial dimensions~\cite{obrien_prb_93_2016}. Doing so for the hypernonagon lattice at hand leads us to the assignment of
$x, y, z$-bonds as illustrated in Fig.~\ref{fig:fig_1}. Note that (up to permutations) there is exactly one such assignment of bonds that
does not break any of the lattice symmetries discussed above. \footnote{Mirror symmetry is, in fact, only unbroken on the line
$J_x = J_y$.}

Since we are particularly interested in the $\mathbb{Z}_2$ gauge physics of this 3D Kitaev model, we readily define an elementary 
$\mathbb{Z}_2$ flux for every plaquette M$_p$ in  analogy to Kitaev models on other lattice geometries as
\begin{equation}
W_p = \prod_{\langle i,j \rangle_\gamma \in {\rm M}_p} \sigma_i^\gamma \sigma_j^\gamma \,.
\label{eq:plaquette}
\end{equation}
Here the product is taken for all the bonds comprising the plaquette in a clockwise manner if viewed from the inside of the orange
hexahedron to which the plaquette belongs. All $W_p$ commute with one another and with the Hamiltonian, and $W_p^2 = -1$ for all $p$.
Hence, $W_p$ is a conserved quantity taking eigenvalues~\footnote{In
Ref.~\onlinecite{kato_prb_96_2017}, the authors defined the $\mathbb{Z}_2$ flux in a different manner by multiplying $-{\mathrm i}$ to
$W_p$ defined there for extracting only a sign in front of ${\rm i}$ of $W_p = \pm {\mathrm i}$.}
 of either $+{\mathrm i}$ or $-{\mathrm i}$. By using this fact, the Hamiltonian can
be block diagonalized, and the Hilbert space of each block (flux sector) is labeled by its flux configuration
$\{W_p\} = \{W_1, W_2, \cdots \}$.

In the present case of the hypernonagon lattice, the $\mathbb{Z}_2$ fluxes $W_p$ have two important features. One comes from the 
dimensionality of the lattice structure. As first discussed for the hyperhoneycomb lattice~\cite{mandal_prb_79_2009}, $\{W_p\}$ is subject
to local and global constraints in three dimensions. In particular, the local constraint, which enforces the product of $W_p$ on any closed
volume to be $+1$, allows excitations by flipping $W_p$ only in the form of closed loops. This leads to a topological finite-$T$ phase
transition between the high-$T$ paramagnet and the low-$T$ quantum spin liquid driven by a proliferation of excited
loops~\cite{nasu_prl_113_2014}. This situation is common to any 3D Kitaev models; indeed, similar finite-$T$ phase transitions were found
on other 3D lattices~\cite{mishchenko_prb_96_2017,arXiv:1901.05283}. In our hypernonagon case, the local constraint is enforced on each
hexahedron in Fig.~\ref{fig:fig_2}: the product of $W_p$ on the eight vertices is always equal to unity in all orange and purple
hexahedra~\cite{kato_prb_96_2017}.

The other important feature of $W_p$ is that it is composed of an odd number of sites for the hypernonagon lattice at hand. When the fluxes
are defined for odd length loops, a pair of flux assignments with opposite signs of $W_p$ form a time-reversal
pair~\cite{kitaev_ann_phys_321_2006}. This allows for the possibility of CSLs where time-reversal symmetry is broken spontaneously by
selecting a particular configuration of $W_p$. Indeed, on the triangle-honeycomb lattice, which consists of 3 and 12-site loops, the ground
state of the Kitaev model was found to be a CSL~\cite{yao_prl_99_2007}. Furthermore, a finite-$T$ phase transition was found from the
high-$T$ paramagnet to the low-$T$ CSL~\cite{nasu_prl_115_2015}.

Thus, our hypernonagon Kitaev model has at least two interesting possibilities for finite-$T$ phase transitions: one is a topological
transition originating from the local constraints on $\{W_p\}$ and the other is the spontaneous breaking of time-reversal symmetry. Such
possibilities were examined in the limits of strong anisotropy in the exchange interactions where finite-$T$ transitions to two different
types of CSLs were found for the effective models. Both transitions have been found to be first order, with the simultaneous occurrence of
the topological change associated with the ordering of the $\mathbb{Z}_2$ gauge structure and the concurrent breaking of a point-group
symmetry of the lattice by nonuniform $\mathbb{Z}_2$ flux ordering~\cite{kato_prb_96_2017}.

%
\section{Methods}
\label{sec:methods}
%

To analyze the hypernonagon Kitaev model introduced in the previous section, we primarily employ numerical approaches to capture the
$\mathbb{Z}_2$ gauge physics of the model: QMC simulations to access the finite-$T$ thermodynamics and variational calculations to map out
the entire zero-$T$ phase diagram. These numerical approaches are complemented by analytical calculations of the Majorana band structures,
which employ various tricks to identify their inherent topological features. In the following, we will briefly outline the relevant details
of all three approaches.

%
\subsection{Quantum Monte Carlo simulation at finite temperatures}
\label{sec:method_qmc}
%

The QMC simulations of the Kitaev model in Eq.~\eqref{eq:kitaev_hamiltonian} are based on a Majorana fermion representation of the
spins~\cite{nasu_prl_113_2014}. Such a representation can be introduced via a Jordan-Wigner transformation on ``chains'' consisting of two
types of bonds (e.g., the $y$ and $z$ bonds)~\cite{chen_prb_76_2007,feng_prl_98_2007,chen_j_phys_a_41_2008}, followed by a rewriting in
terms of the two types of Majorana fermions $c$ and $\bar{c}$, as
\begin{align}
\mathcal{H} = \mathrm{i}J_y\sum_{\langle i,j \rangle_y}           c_i c_j
            - \mathrm{i}J_z\sum_{\langle i,j \rangle_z}           c_i c_j
            - \mathrm{i}J_x\sum_{\langle i,j \rangle_x} \eta_{ij} c_i c_j \,,
\label{eq:majorana_hamiltonian}
\end{align}
where the sums are taken for $i < j$, and $\eta_{ij} = {\mathrm i} \bar{c}_i \bar{c}_j$. Thus, the Majorana fermions described by $c$ have
an itinerant nature, while those described by $\bar{c}$ are localized on the $x$ bonds. Here, $\eta_{ij}$ is also a local conserved
quantity taking eigenvalue either $+1$ or $-1$, which is related to $W_p$ in Eq.~\eqref{eq:plaquette} as $W_p = {\mathrm i}
\prod_{\langle i,j \rangle_x \in p} \eta_{ij} \label{eq:majorana_plaquette}$, where the product is taken for all the $x$ bonds included in
the 9-site loop. This Majorana fermion representation enables us to perform sign-free QMC simulations by sampling configurations of the
$\mathbb{Z}_2$ variables $\{\eta_{ij}\}$.

In the Monte Carlo (MC) sampling, we perform single flip updates of the $\mathbb{Z}_2$ variables $\{\eta_{ij}\}$ as well as replica
exchange~\cite{hukushima_jpsj_65_1996} updates with respect to $T$. In the single flip updates, we compute the Boltzmann weight
difference caused by a local sign flip of $\eta_{ij}$ by employing the Green-function-based kernel polynomial method introduced in
Refs.~\onlinecite{weisse_prl_102_2009,weisse_rev_mod_phys_78_2006} and applied to a Kitaev model in
Ref.~\onlinecite{mishchenko_prb_96_2017}. The computational complexity of a single update is thereby reduced to $\mathcal{O}(N)$ (where $N$
is the number of sites), which is a big advantage compared to the $\mathcal{O}(N^3)$ scaling of a naive approach based on exact
diagonalizations of the Hamiltonian. In addition, replica exchange updates are introduced to prevent the system from freezing at low $T$.
To enhance its overall efficiency, we adopt a feedback optimization~\cite{katzgraber_jsm_03018_2006} of the $T$ set for parallel tempering,
which significantly speeds up thermal equilibration~\footnote{For the purpose of better accuracy we used the exact diagonalization for
replica exchange updates and for measurements. Eventually, complexity of the entire simulation becomes $\mathcal{O}(N^3)$.}.

In the setup of our QMC simulations, we consider two clusters with a total of $N = 12 \times L^3$ sites: one has the linear dimension
$L = 4$ ($N = 768$) and the other has $L = 6$ ($N = 2592$). We employ periodic boundary conditions in all spatial directions in order
to minimize finite-size effects in our simulations~\footnote{We note that both open and periodic boundary conditions were adopted in the
previous studies~\cite{nasu_prl_113_2014,mishchenko_prb_96_2017,dwivedi_prb_98_2018}. The results for different boundary conditions are
expected to converge in the thermodynamic limit.}. We optimize the $T$ set by the feedback optimization
technique~\cite{katzgraber_jsm_03018_2006}, where we initially perform $3000$ MC sweeps for sampling after $3000$ MC sweeps for
thermalization, and then add $1000$ MC sweeps to both sampling and thermalization after each successive optimization iteration. We perform
seven and four steps of the feedback optimization for the $L = 4$ and $L = 6$ systems, respectively, where, for the optimization of the
$L = 6$ system, we start from the $T$ set obtained from the fifth step for the $L = 4$ system. We present the results at the sixth step
($4800$ MC samples) and the seventh step ($4000$ MC samples) for the $L = 4$ system, and at the third step ($1600$ MC samples) and the
fourth step ($4800$ MC samples) for the $L = 6$ system. The measurements are performed after every MC sweep~\footnote{All the data are
presented with $3\sigma_\mathrm{SE}$ errorbars, where $\sigma_\mathrm{SE}$ stands for the standard error}.

%
\subsection{Variational calculations for the ground state}
\label{sec:variational_calculation}
%

Probing several different configurations of the $\mathbb{Z}_2$ fluxes (see Sec.~\ref{sec:zero-t}), we obtain the corresponding energies by
exact diagonalization of the Hamiltonian in Eq.~\eqref{eq:majorana_hamiltonian}. From the comparison, we determine the variational ground
state as the state with the lowest energy for a given set of the exchange couplings, $J_x$, $J_y$, and $J_z$. The energy of a particular
flux configuration $\{W_p\}$ is calculated by constructing the $\{\eta_{ij}\}$ configuration that reproduces the given $\{W_p\}$ and
diagonalizing the Fourier transform of the Hamiltonian in Eq.~\eqref{eq:majorana_hamiltonian} for the obtained $\{\eta_{ij}\}$. We note
that $\{W_p\}$ for all the variational states are reproduced by a particular $\{\eta_{ij}\}$ within the $L = 2$ unit cell. In these
calculations, we consider a system with linear system size $L = 48$.

%
\subsection{Determination of the Majorana band structures}
\label{sec:method_majorana}
%

Given a variational flux configuration for a specific choice of the exchange couplings, the fermionic ground state is obtained by computing
the corresponding spectrum of the itinerant Majorana fermions. In order to efficiently determine whether the fermionic quasiparticles are
gapped or gapless at a given point in the phase diagram, the  Berry curvature is integrated over 2D planes~\cite{FukuiJPS2005} which cut
through the 3D Brillouin zone, i.e., by fixing one component of the momentum and integrating over the other two.

For those planes on which the fermionic spectrum is gapped, the result is a quantized Chern number. In a Weyl spin
liquid~\cite{HermannsPRL2015} phase, the Chern number jumps discontinuously as the plane passes through a Weyl node by an amount equal to
the charge of that Weyl node -- thereby allowing for a robust identification of the location of Weyl nodes in momentum space. In general,
the Berry curvature is ill-defined for a plane on which the fermionic spectrum is gapless. In the case that there is a one-dimensional (1D)
(line) or 2D (plane) nodal manifold, the result is a range of momentum values for which the Berry flux is ill-defined and typically
fluctuates wildly.

With this information, it may be determined whether a given point in the phase diagram corresponds to a gapped or gapless fermionic
spectrum. If the Berry flux is  vanishing everywhere, the fermions must be gapped throughout the entire Brillouin zone. However, if at any
point the Berry flux is \textit{nonzero} -- whether it takes a nonvanishing quantized Chern number or fluctuates wildly -- the fermionic
spectral gap must close somewhere. For the regions of the phase diagram where the gap closing occurs at a high-symmetry point, the
resolution of the phase diagram data can be further increased by checking whether or not the gap closes at that point for a given choice of
couplings.

Alongside this analysis, the projective symmetry group~\cite{WenPRB2002} is utilized to deduce general constraints on the gapless fermionic
excitations, thereby explaining the distinct stable topology of the nodal manifold of the itinerant Majorana fermions appearing in the
different CSL phases. The precise projective representation of physical symmetries of the model when written in the Majorana fermion
representation determines the character of the nodal manifold~\cite{obrien_prb_93_2016, Yamada2017_crystalline}. In particular, the
projective symmetry group may enforce the stability of either fully 2D Fermi surfaces, nodal lines, Dirac cones or topological Weyl nodes.

%
\section{Thermodynamics}
\label{sec:finite-t}
%

We start our presentation of results by first discussing the thermodynamics of the hypernonagon Kitaev model.

%
\subsection{Thermal fractionalization}
\label{sec:fractionalization}
%

\begin{figure}[t]
	\includegraphics[width=\columnwidth,clip]{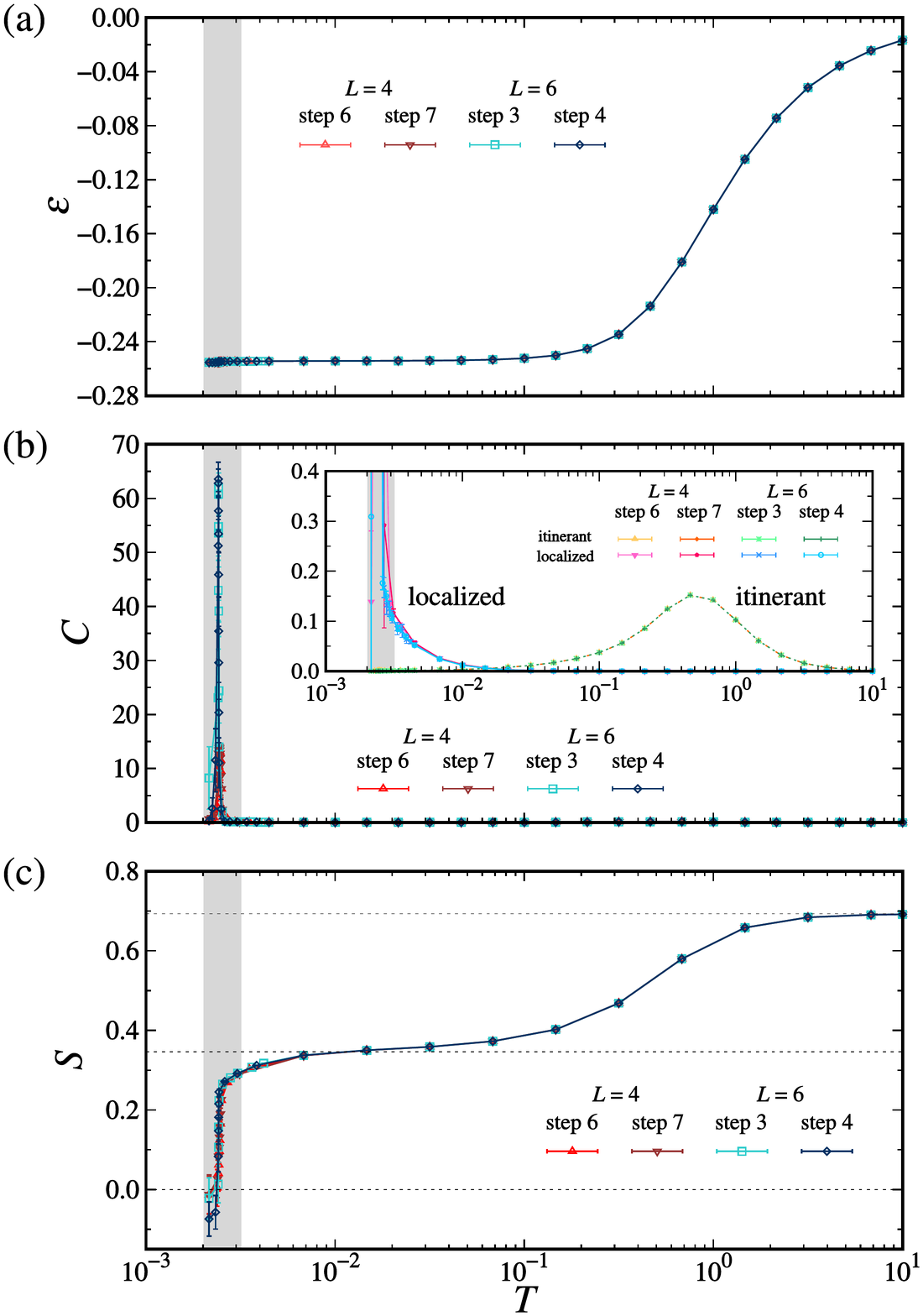}
	\caption{
		Thermodynamics of the hypernonagon Kitaev model obtained by QMC simulations for the isotropic exchange interactions
		$J_x = J_y = J_z = 1/3$. Shown are the $T$ dependence of (a) the internal energy $\varepsilon$, (b) the specific heat $C$, and (c) the
		entropy $S$ per site at the last two optimization steps. The inset of (b) displays the decomposition into the contributions from the
		itinerant Majorana fermions and the localized $\mathbb{Z}_2$ fluxes. The data are calculated for the system sizes with $L = 4$ and
		$L = 6$. The gray area indicates the $T$ range plotted in Fig.~\ref{fig:fig_4}. The dashed horizontal lines in (c) indicate $0$,
		$\frac{1}{2}\ln 2$, and $\ln 2$.
	}
	\label{fig:fig_3}
\end{figure}

Figure~\ref{fig:fig_3} summarizes our QMC results for the isotropic Kitaev model with equal couplings $J_x = J_y = J_z$, normalized to 
$J_x + J_y + J_z = 1$. It shows (a) the internal energy $\varepsilon$, (b) the specific heat $C$, and (c) the entropy $S$ per site as
functions of $T$ at the last two optimization steps for each system size as described in Sec.~\ref{sec:method_qmc}. In this resolution, the
two results appear to overlap; we will discuss the convergence in Sec.~\ref{sec:transition}. Similar to the previous works on other
lattices~\cite{nasu_prl_113_2014,nasu_prb_92_2015,mishchenko_prb_96_2017}, the specific heat $C$ of the Kitaev model on the hypernonagon
lattice exhibits two peaks as shown in Fig.~\ref{fig:fig_3}(b), corresponding to the steep changes of the energy $\varepsilon$ plotted in
Fig.~\ref{fig:fig_3}(a) [see the low-$T$ part in Fig.~\ref{fig:fig_4}(a)]. One is a broad peak at $T \approx 0.5$ and the other is a sharp
peak at $T \approx 0.0024$. The high-$T$ peak comes from the itinerant Majorana fermions $c$, while the low-$T$ peak is due to the
$\mathbb{Z}_2$ fluxes, which are related to the localized Majorana fermions $\bar{c}$~\cite{nasu_prl_113_2014,nasu_prb_92_2015}. This is
clearly demonstrated by a decomposition of the specific heat $C$ into each contribution, as shown in
Fig.~\ref{fig:fig_3}(b)~\cite{dwivedi_prb_98_2018}. Figure~\ref{fig:fig_3}(c) demonstrates that half of the total entropy $S$,
$\frac{1}{2}\ln 2 \approx 0.347$, is released at each peak of the specific heat $C$~\footnote{The entropy $S$ is calculated by numerical
integration of the internal energy $\varepsilon$ from high $T$. We note a small negative value of $S$ at the lowest $T$ in some cases,
which is presumably due to the slow convergence in the vicinity of the phase transition.}. This successive release of the entropy is a
common feature to all Kitaev models, indicative of the thermal fractionalization of spins~\cite{nasu_prb_92_2015}.

%
\subsection{Finite-temperature phase transition}
\label{sec:transition}
%

\begin{figure}[tb]
	\includegraphics[width=\columnwidth,clip]{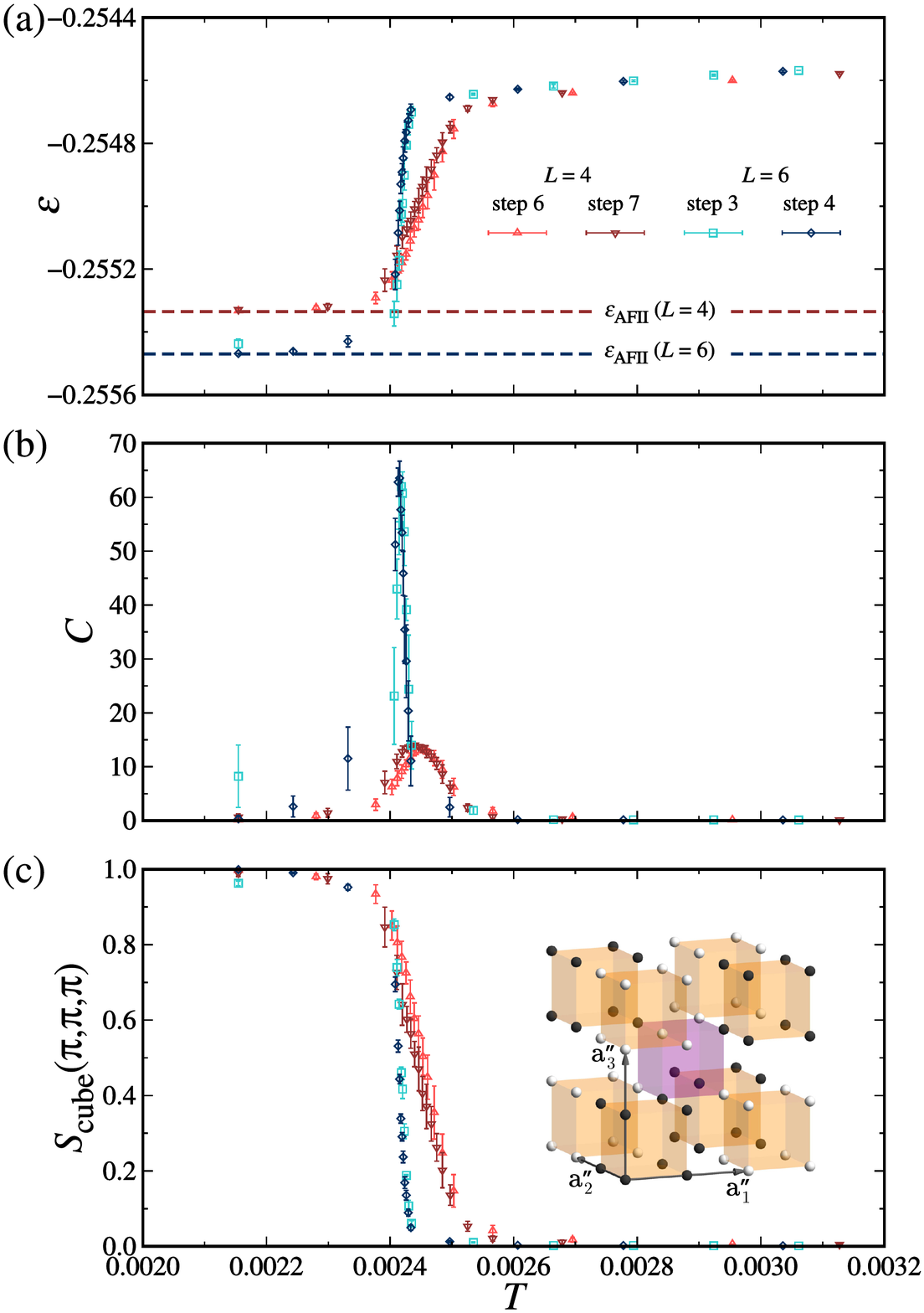}
	\caption{
		Low-$T$ phase transition. Shown are the $T$ dependence of (a) the internal energy $\varepsilon$ and (b) the specific heat $C$ as given
		in Figs.~\ref{fig:fig_3}(a) and \ref{fig:fig_3}(b), respectively, but plotted on a linear $T$ scale, zooming in the low-$T$ regime (the
		gray areas in Fig.~\ref{fig:fig_3}). (c) The order parameter for the $\mathbb{Z}_2$ flux ordering, $S_{\rm cube}(\textbf{Q})$ at
		$\textbf{Q} = (\pi,\pi,\pi)$ at the last two optimization steps. The inset of (c) indicates the flux configuration in the low-$T$ CSL
		phase, where black and white spheres represent fluxes with $W_p = +{\rm i}$ and $-{\rm i}$, respectively. $\varepsilon_{\rm AFII}$ in
		(a) indicates the ground state energy for each system size. See the main text for details.
	}
	\label{fig:fig_4}
\end{figure}

In the following, we focus on the low-$T$ region where the specific heat $C$ shows a sharp peak (gray area in Fig.~\ref{fig:fig_3}).
Figure~\ref{fig:fig_4} displays the low-$T$ QMC data. The internal energy $\varepsilon$ suddenly drops at $T \approx 0.0024$, as shown in
Fig.~\ref{fig:fig_4}(a). The energy change becomes sharper for $L = 6$ than $L = 4$. Correspondingly, the specific heat $C$ shows a sharp
peak, as shown in Fig.~\ref{fig:fig_4}(b). We find that the peak height (width) becomes higher (narrower) for larger $L$, indicating a
phase transition~\footnote{We note that the convergence of the data close to the phase transition is extremely slow; there still remain
discrepancies between the last two optimization steps. Nonetheless, our QMC results are sufficient for the discussion presented in the
main text.}.

\begin{figure}[tb]
	\includegraphics[width=\columnwidth,clip]{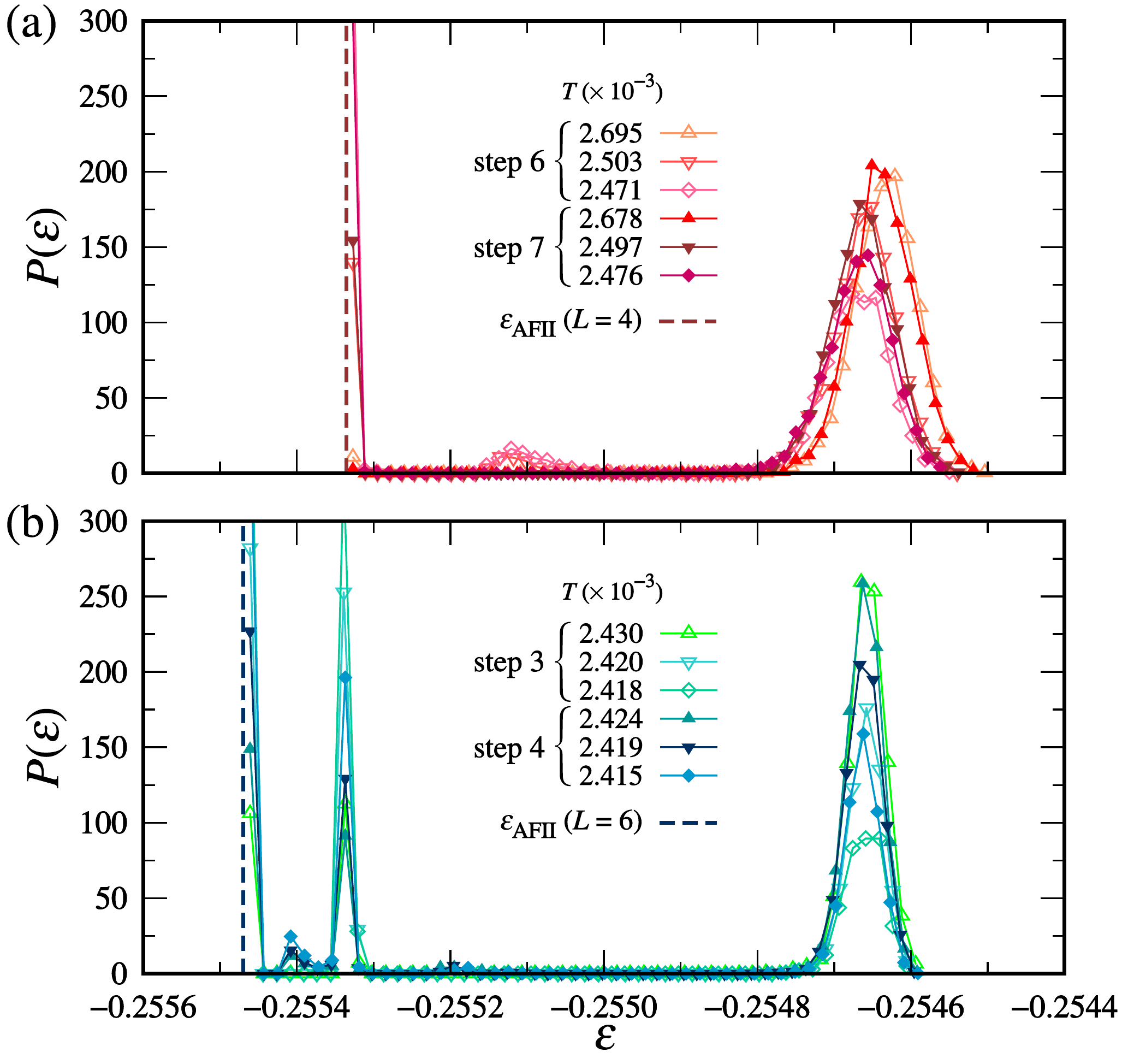}
	\caption{
		First-order nature of the thermal phase transition. Histograms of the internal energy $\varepsilon$, $P(\varepsilon)$, in the vicinity
		of the transition temperature $T_c = 0.00244(4)$ obtained by QMC simulations for (a) $L = 4$ and (b) $L = 6$ at the last two
		optimization steps. $\varepsilon_{\rm AFII}$ is the internal energy calculated for the $W_p$ configuration in the AFII phase for each
		system size. See the text for details.
	}
	\label{fig:fig_5}
\end{figure}

In order to clarify the nature of the phase transition, we calculate the histogram of the internal energy $\varepsilon$, $P(\varepsilon)$.
The results are shown in Fig.~\ref{fig:fig_5}. We find that the energy distribution bifurcates for both system sizes, while it has an
additional structure in the low-energy part (we return to this point later). The bifurcation strongly suggests a first-order phase
transition, which is consistent with the size dependence of the specific heat $C$, namely, the peak height increases roughly proportional
to the system size as expected for a first-order phase transition.

\begin{figure*}[tb]
	\includegraphics[width=\textwidth]{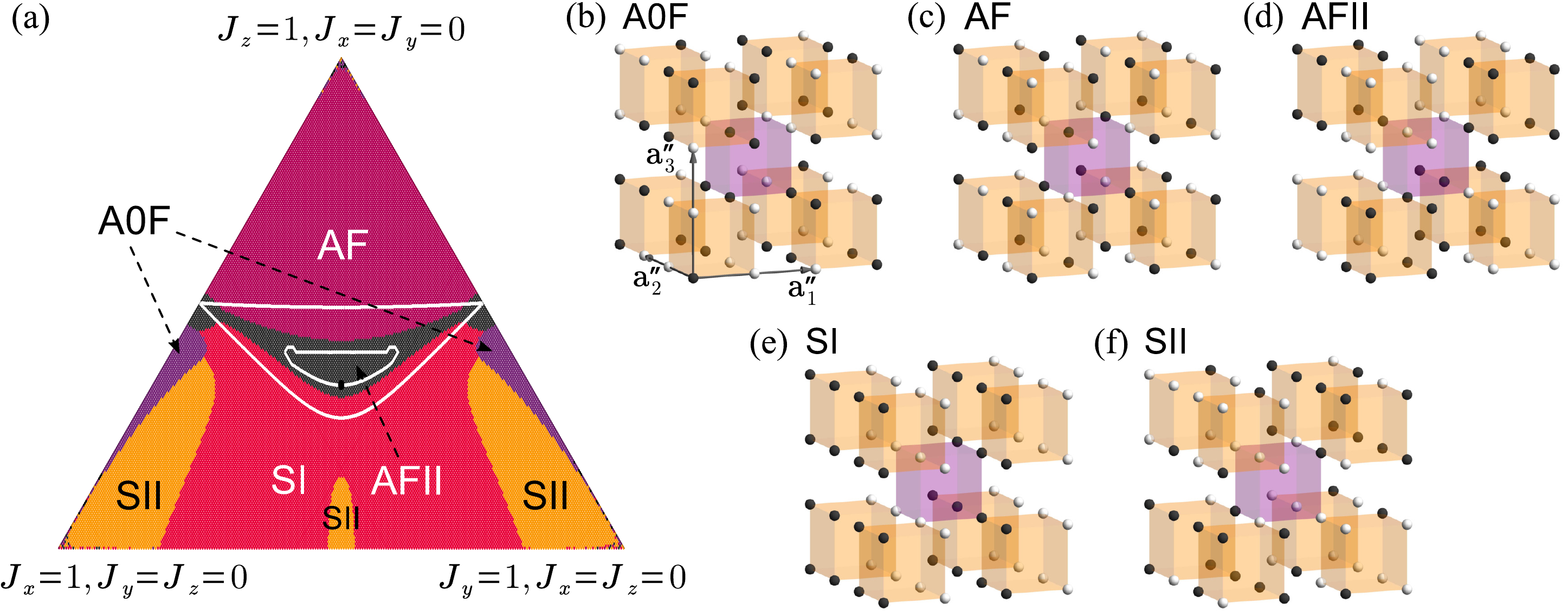}
	\caption{\label{fig:fig_6}
		Ground state of the hypernonagon Kitaev model. (a) The ground state phase diagram in the parameter space $J_x + J_y + J_z = 1$
		($J_{\gamma} \geq 0$) obtained by variational calculations. Each corner corresponds to the anisotropic limits where one of
		$J_{\gamma} = 1$ and the other two are zero. The black dot at the center represents the isotropic case $J_x = J_y = J_z = 1/3$. The
		white lines represent gapped/gapless boundaries with respect to the itinerant Majorana spectrum; the system has gapless excitations in
		the area sandwiched by the two closed white lines. Five distinct phases are found: A0F (purple), AF (magenta), AFII (gray), SI (red),
		and SII (yellow). The flux configurations are shown in (b)-(f) for each phase. The black and white spheres represent fluxes with
		$W_p = +{\mathrm i}$ and $-{\mathrm i}$, respectively. Since all the variational states have the same energy in the corners of the
		phase diagram, numerical noise may be seen close to them.
	}
\end{figure*}

We find that the finite-$T$ phase transition is caused by ordering of the $\mathbb{Z}_2$ fluxes $W_p$. This can be monitored by the
structure factor of $W_p$, defined in matrix form as
\begin{align}
S^{\alpha\beta}_{\rm cube}(\textbf{k}) = \frac{1}{8^2 L^3}
\langle W_\alpha(\textbf{k}) \cdot \lbrack W_\beta(\textbf{k}) \rbrack^* \rangle \,,
\label{eq:structure_factor}
\end{align}
where
\begin{align}
W_\alpha(\textbf{k}) = \frac{1}{\sqrt{L^3}} \sum_{\textbf{r}} \mathrm{e}^{-\mathrm{i} \textbf{k} \cdot \textbf{r}} W_\alpha(\textbf{r}) \,.
\end{align}
Here, $\textbf{r}$ is the position vector for the orange hexahedral unit cell in Fig.~\ref{fig:fig_2}(b) and $\alpha, \beta \in \{1, 2,
\ldots, 8\}$ label the vertices of the hexahedron. We find that the structure factor develops a Bragg peak at $\textbf{k} = \textbf{Q} =
(\pi, \pi, \pi)$, with all the matrix elements of $S^{\alpha\beta}_{\rm cube}(\textbf{Q})$ being equivalent. Hence, we define the order
parameter $S_{\rm cube}(\textbf{Q}) \equiv \sum_{\alpha,\beta} S^{\alpha\beta}_{\rm cube}(\textbf{Q})$ and plot its value as a function of
$T$ in Fig.~\ref{fig:fig_4}(c). The result indicates that the first-order transition of the system is caused by a spontaneous ordering of
the $\mathbb{Z}_2$ fluxes where all fluxes in a given hexahedral unit have the same sign, while the units themselves form a staggered
pattern. This is illustrated in the inset of Fig.~\ref{fig:fig_4}(c). We call this flux order AFII (see Sec.~\ref{sec:zero-t}). As
mentioned above, the ordering of $\{W_p\}$ corresponds to a spontaneous breaking of time-reversal symmetry. Hence, the transition observed
here is deduced to occur from the high-$T$ paramagnetic phase to the low-$T$ quantum spin liquid phase with broken time-reversal symmetry,
i.e., a CSL phase. In Fig.~\ref{fig:fig_4}(a), we plot the ground state energy $\varepsilon_{\rm AFII}$ for this CSL phase for each $L$.
Our QMC data quickly converge to these values below the transition $T$.

The peak $T$ of the specific heat $C$ does not change much from $L = 4$ to $L = 6$, as shown in Fig.~\ref{fig:fig_4}(b). We estimate the
transition $T$ to be $T_c = 0.00244(4)$ so that both values are included in the error in the last digit, although more precise
determination is necessary by the system-size extrapolation. We do not see any anomaly except at $T_c$ in the $T$ range studied here. This
suggests that the hypernonagon Kitaev model at the isotropic point exhibits a single first-order transition to the CSL, similar to what is
observed in its anisotropic limits~\cite{kato_prb_96_2017}. However, given the peculiar structure of $P(\varepsilon)$ in
Fig.~\ref{fig:fig_5}, the possibility of multiple transitions within a very narrow $T$ range cannot be excluded at present, and needs to be
examined carefully for larger system sizes. This is left for a future study.

In the 3D Kitaev models, the transition temperature $T_c$ is related to the loop tension, which defines the energy gap for a loop
excitation of $W_p$~\cite{kimchi_prb_90_2014,tim_2019}. In our hypernonagon case, we estimate the flux gap, defined by the smallest
excitation of a $W_p$ loop from the ground state, as $\Delta = 0.0339(1)$. The previous studies for other 3D lattices showed that the ratio
$T_c/\Delta$ is in the range of $\approx 0.11$--$0.15$~\cite{nasu_prl_113_2014,mishchenko_prb_96_2017,tim_2019}. In the hypernonagon case,
however, the value of $T_c/\Delta \approx 0.07$ deviates from this trend: $T_c$ is considerably lower than the expected value given
$\Delta$. The origin of the deviation can likely be traced back to the first-order nature of the transition in the hypernonagon lattice,
where multiple symmetries (time-reversal and lattice point group) are broken simultaneously with the gauge ordering transition. In
contrast, the transition for all other 3D Kitaev models appears to be continuous \cite{tim_2019}, indicating that similar physics is at
play as for the  hyperhoneycomb and hyperoctagon cases for which the finite-$T$ transition has been shown to arise from a topological
change of the loop-type excitations~\cite{nasu_prb_89_2014,nasu_prl_113_2014,mishchenko_prb_96_2017}.

%
\section{Ground state phase diagram}
\label{sec:zero-t}
%

Our calculations reveal several low-$T$ flux configurations when probing the phase diagram away from the isotropic coupling point, which
all exhibit nonuniform arrangements of the $\mathbb{Z}_2$ fluxes. The five distinct arrangements are illustrated in
Figs.~\ref{fig:fig_6}(b)-\ref{fig:fig_6}(f), which in the following we will refer to as A0F, AF, AFII, SI, and SII, respectively. Most of
these were previously identified in certain parameter regimes, such as the SI configuration of Fig.~\ref{fig:fig_6}(e) in initial
studies~\cite{obrien_prb_93_2016} of the hypernonagon model by using variational arguments, the A0F configuration of
Fig.~\ref{fig:fig_6}(b) in Ref.~\onlinecite{kato_prb_96_2017}, the SII configuration of Fig.~\ref{fig:fig_6}(f) in the strong $J_x$ or
$J_y$ limits~\cite{kato_prb_96_2017} with a subextensive degeneracy, and the AF configuration in the strong $J_z$
limit~\cite{kato_prb_96_2017}. Only the AFII phase of Fig.~\ref{fig:fig_6}(d) has hitherto remained unnoticed.

Using these five distinct ordering patterns as input for the variational calculations described in Sec.~\ref{sec:variational_calculation},
we can map out the entire zero-$T$ phase diagram of the $\mathbb{Z}_2$  flux sector for arbitrary Kitaev couplings. The result is
summarized in Fig.~\ref{fig:fig_6}(a) for the plane of $J_x + J_y + J_z = 1$ with $J_\gamma \geq 0$ (with the phase diagram remaining the
same when flipping the signs of all $J_\gamma$). In the phase diagram we find the AFII phase (gray area) around the isotropic point (black
dot), consistent with our QMC results. When increasing $J_z$ (moving upwards in the phase diagram), the AFII phase gives way to an AF phase
(magenta). On the other hand, when decreasing $J_z$ we find the SI phase (red), and finally the SII phase (yellow) at the bottom center as
well as near the two bottom corners of the phase diagram. We also find narrow regions of the A0F phase (purple) near the edges of $J_x = 0$
or $J_y = 0$. We note that our results around the three corners are indeed consistent with ealier results~\cite{kato_prb_96_2017}, derived
perturbatively for the three anisotropic limits.

In addition to the phase boundaries arising from the $\mathbb{Z}_2$ flux ordering, we also determine the boundaries that arise from the
transition between gapless and gapped band structures of the itinerant Majorana fermions -- as indicated by the white lines in the phase
diagram of Fig.~\ref{fig:fig_6}(a). We find that in general these boundaries do not precisely coincide with the phase boundaries between
the different flux patterns, i.e., for a given flux pattern there are regions where the Majorana spectrum is either gapped or gapless. In
particular, the extent of the gapless region of the Majorana band structure corresponds to the region sandwiched by the two closed white
lines in the phase diagram of Fig.~\ref{fig:fig_6}(a). It is thus that the SI, AF, and AFII phases all have both gapped and gapless
Majorana spectra, whereas for the A0F and SII phases the Majorana spectrum remains fully gapped for all parameters. We note that the
isotropic point is right on the gapped/gapless boundary. 

In the following sections, the nature of the gapless regions are discussed for the SI, AF, and AFII phases in further detail. As the
lattice symmetries relevant to this discussion are the $C_3$-rotational, mirror, and inversion symmetries, the corresponding symmetry
information for the different CSL flux phases is summarized in Table~\ref{table:PhaseSymmetries}. Among the flux phases which host gapless
fermions, only the SI phase breaks the rotational symmetry and two out of the three mirror symmetries. While the SI and AF phases respect
inversion symmetry for \textit{all} inversion centers, the AFII phase breaks inversion symmetry for the inversion centers corresponding to
the orange hexahedra [see Fig.~\ref{fig:fig_6}(d)]. In addition as discussed in Sec.~\ref{sec:flux}, by virtue of the hypernonagon lattice
being non-bipartite, all flux configurations necessarily break time-reversal symmetry. For the AFII phase, however, the combination of
time-reversal with the broken inversion symmetry operation is seen to yield a symmetry of the Hamiltonian. 

\begin{table}[b]
	\centering
	\begin{tabular}{ccccccc}
		\hline
		Flux phase & \qquad	&	$C_3$	&	\qquad & Mirror planes & \qquad	&	Inversion (purple / orange)	\\ \hline
		A0F	       &			  &	Yes		&			   &	3				     &			  &	Yes / No	            			\\
		AF         &			  &	Yes		&			   &	3				     &			  &	Yes / Yes				            \\
		AFII       &			  &	Yes		&			   &	3				     &			  &	Yes / No				            \\
		SI         &			  &	No		&			   &	1				     &			  &	Yes / Yes				            \\
		SII        &			  &	No		&			   &	1				     &			  &	No / No
	\end{tabular}
	\caption{
	Overview of lattice symmetries for the different CSL phases. The relevant symmetries include $C_3$-rotational symmetry, the number of
	remaining mirror planes (for $J_x = J_y$) and whether inversion symmetries (corresponding to the purple and orange hexahedra) are
	present.
	}
	\label{table:PhaseSymmetries}
\end{table}

The physical inversion symmetry for all the CSL phases discussed here is represented projectively as
\begin{equation}
H(\textbf{k})
= G^\dagger_{\mathcal{P}}~U^{\phantom\dagger}_{\mathcal{P}}~H(-\textbf{k})~U^\dagger_{\mathcal{P}}~G^{\phantom\dagger}_{\mathcal{P}},
\label{eq:ProjectiveInversion}
\end{equation}
where $U_{\mathcal{P}}$ is the matrix representation of the inversion operator acting on the unit cell and $G_\mathcal{P}$ is the
associated $\mathbb{Z}_2$ gauge transformation. Note that for all the phases at hand, inversion symmetry is implemented trivially, i.e.,
with a uniform gauge transformation $G_\mathcal{P}$. A nonuniform gauge transformation necessitates a translation of $\mathbf k$ by half a
reciprocal lattice vector on the right side of Eq.~\eqref{eq:ProjectiveInversion}~\cite{obrien_prb_93_2016}. The relevant energy relations
are, thus, given by
\begin{equation}
E_{\alpha}(\textbf{k}) = -E_{\beta}(-\textbf{k}) \quad {\rm and} \quad E_{\alpha}(\textbf{k}) = E_{\gamma}(-\textbf{k}),
\label{eq:EnergyRelation}
\end{equation}
due to particle-hole and inversion symmetry, respectively. Here $\alpha$, $\beta$, and $\gamma$ denote band indices. 

%
\subsection{SI phase}
\label{section:chapter06_SIPhase}
%

\begin{figure}[t]
	\centering
	\includegraphics[width=\columnwidth,clip]{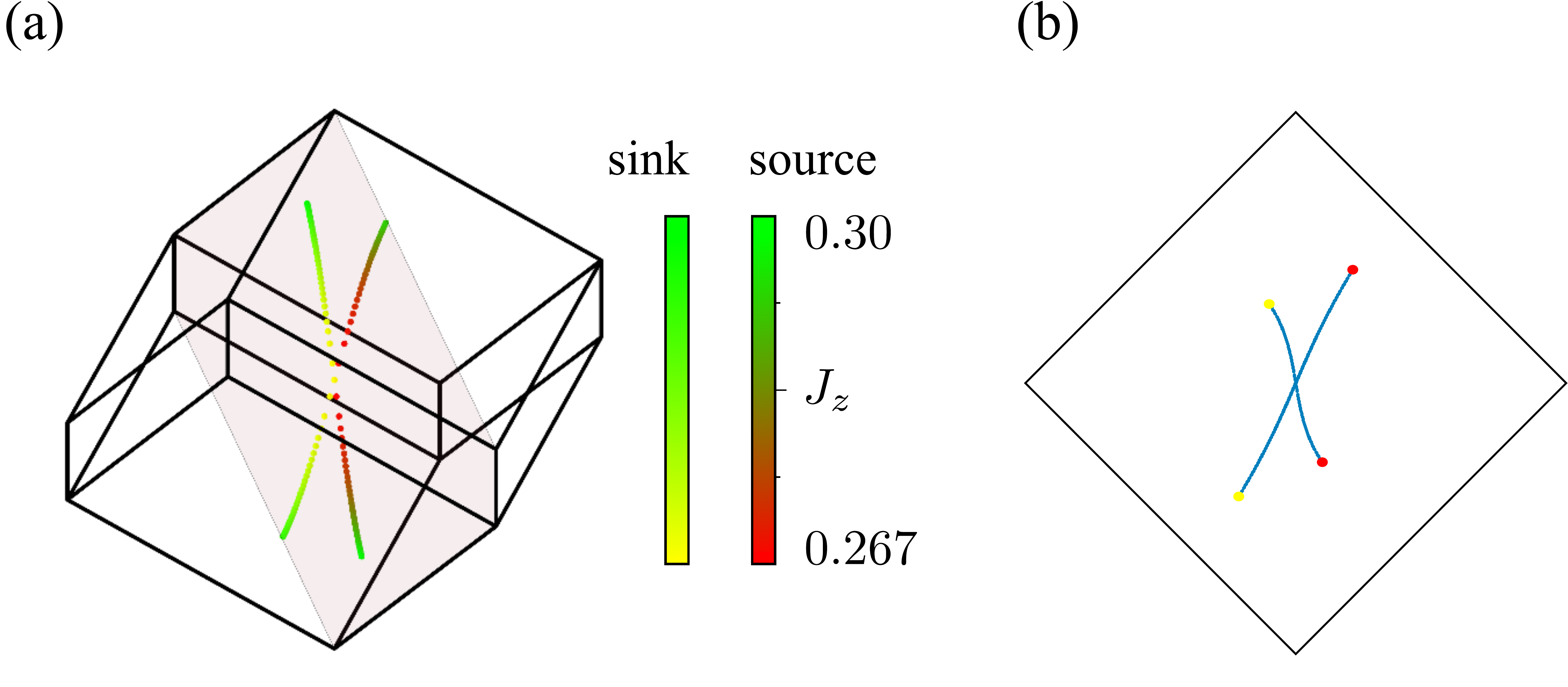}
	\caption{
		Gapless SI phase. (a) Evolution of Weyl nodes for coupling constants $0.267 \leq J_z \leq 0.30$ with $J_x = J_y = (1 - J_z)/2$. The
		sink and source indicate the negatively- and positively-charged Weyl nodes, respectively. The shaded region indicates the remaining
		mirror plane in the SI phase with $J_x = J_y$. (b) Corresponding Fermi arcs on the 001-surface Brillouin zone for $(J_x, J_y, J_z) =
		(0.36, 0.36, 0.28)$.
	}
	\label{fig:SIPanel}
\end{figure}

In the SI phase, the presence of trivially-implemented projective inversion symmetries prohibits the formation of stable Fermi surfaces due
to the corresponding energy relation in Eq.~\eqref{eq:EnergyRelation}~\cite{hermanns_prb_89_2014}. Furthermore, the absence of
time-reversal symmetry prohibits the formation of stable nodal lines seen in some other Kitaev spin liquids, e.g., on the hyperhoneycomb
lattice~\cite{obrien_prb_93_2016}, where such nodal lines are stabilized by a 1D winding number~\cite{BurkovPRB2011,ZhaoPRL2013,
MatsuuraNJP2013} associated with sublattice symmetry, i.e. a combination of  trivially-implemented  time-reversal and particle-hole
symmetry. However, the combination of particle-hole and inversion symmetries allows for the presence of topologically protected Weyl nodes
pinned to zero energy.

Indeed, diagonalizing the concrete Hamiltonian in the SI phase reveals an extended gapless Weyl spin liquid phase. Restricting the exchange
couplings to the line $J_x = J_y = (1 - J_z)/2$, the gapless portion of the SI phase runs from $J_z \approx 0.267$, where the gap closes,
to $J_z \approx 0.3$, where the ground state switches from the SI phase to the AFII phase. For $J_z \approx 0.267$, two positively-charged
and two negatively-charged Weyl nodes simultaneously appear at the $\Gamma$ point of the Brillouin zone. As $J_z$ is further increased, the
Weyl nodes split apart. For $J_x = J_y$, the Weyl nodes are pinned to the single mirror plane which is \textit{not} broken in the SI phase.

Figure~\ref{fig:SIPanel}(a) shows the evolution of the Weyl nodes in the 3D Brillouin zone as the exchange couplings are varied for
$0.276 \leq J_z \leq 0.30$ with $J_x = J_y = (1 - J_z)/2$. The trajectory of negatively-charged Weyl nodes changes color from yellow to
green as $J_z$ is increased, whereas the trajectory of positively-charged Weyl nodes changes from red to green. The corresponding Fermi
arcs at the 001-surface Brillouin zone for $(J_x, J_y, J_z) = (0.36, 0.36, 0.28)$ are plotted in Fig.~\ref{fig:SIPanel}(b).

%
\subsection{AF phase}
\label{section:chapter06_AFPhase}
%

In the AF phase, as for the SI phase discussed above, the presence of trivially-implemented projective inversion symmetries prohibits the
formation of stable Fermi surfaces. Furthermore, the absence of time-reversal symmetry prevents the formation of nodal lines protected by
sublattice symmetry as mentioned above.  However, the combination of particle-hole and inversion symmetries allows for the presence of
topologically protected Weyl nodes pinned to zero energy.

Indeed, diagonalizing the concrete Hamiltonian in the AF phase reveals an extended region with Weyl nodes. Restricting the exchange
couplings to the line $J_x = J_y = (1 - J_z)/2$, the gapless portion of the AF phase runs from $J_z \approx 0.42$, where the ground state
switches from the AFII phase to the AF phase, to $J_z \approx 0.491$, where the Weyl nodes gap out. For $0.42 < J_z \lesssim 0.491$, two
oppositely-charged Weyl nodes are pinned to the $C_3$-invariant axis. As $J_z$ is increased, the two Weyl nodes move towards each other
until they eventually meet at the $\Gamma$ point for $J_z \approx 0.491$ and mutually annihilate.

\begin{figure}[tb]
	\centering
	\includegraphics[width=\columnwidth,clip]{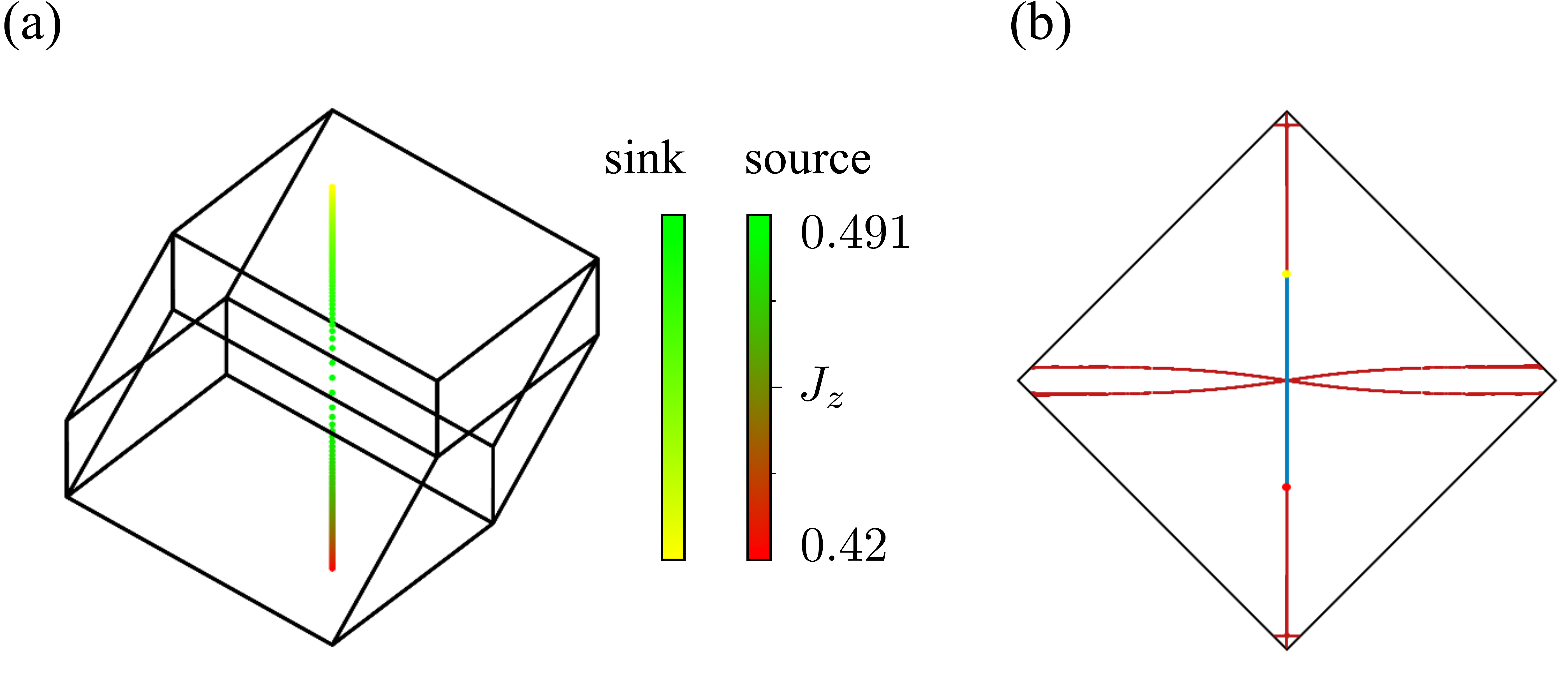}
	\caption{
		Gapless AF phase. (a) Evolution of Weyl nodes for coupling constants $0.42 \leq J_z \leq 0.491$ with $J_x = J_y = (1 - J_z)/2$. The
		sink and source indicate the negatively- and positively-charged Weyl nodes, respectively. (b) Corresponding Fermi arc in the
		001-surface Brillouin zone for $(J_x, J_y, J_z) = (0.27, 0.27, 0.46)$ denoted in blue. Additional surface states are marked in red and
		are discussed in Appendix~\ref{sec:appendix2}.
	}
	\label{fig:AFPanel1}
\end{figure}

Figure~\ref{fig:AFPanel1}(a) shows the evolution of the Weyl nodes in the 3D Brillouin zone as the exchange couplings are varied for
$0.42 \leq J_z \leq 0.491$ with $J_x = J_y = (1 - J_z)/2$. The trajectory of negatively-charged Weyl nodes (sinks) changes color from
yellow to green as $J_z$ is increased, whereas the trajectory of positively-charged Weyl nodes (sources) changes from red to green. The
corresponding Fermi arc at the 001-surface of the Brillouin zone for $(J_x, J_y, J_z) = (0.27, 0.27, 0.46)$ is plotted in blue in
Fig.~\ref{fig:AFPanel1}(b).

In addition to the Fermi arc which terminates at the projection of the Weyl nodes in the surface Brillouin zone, there appear additional
line-like surface states which form incontractible loops. These can be seen as remnants of Fermi arcs due to the Weyl nodes which are
gapped out in this part of the phase diagram. For a thorough discussion of these additional surface states, we refer to
Appendix~\ref{sec:appendix2}.

%
\subsection{AFII phase}
\label{section:chapter06_AFIIPhase}
%

\begin{figure}[b]
	\centering
	\includegraphics[width=\columnwidth,clip]{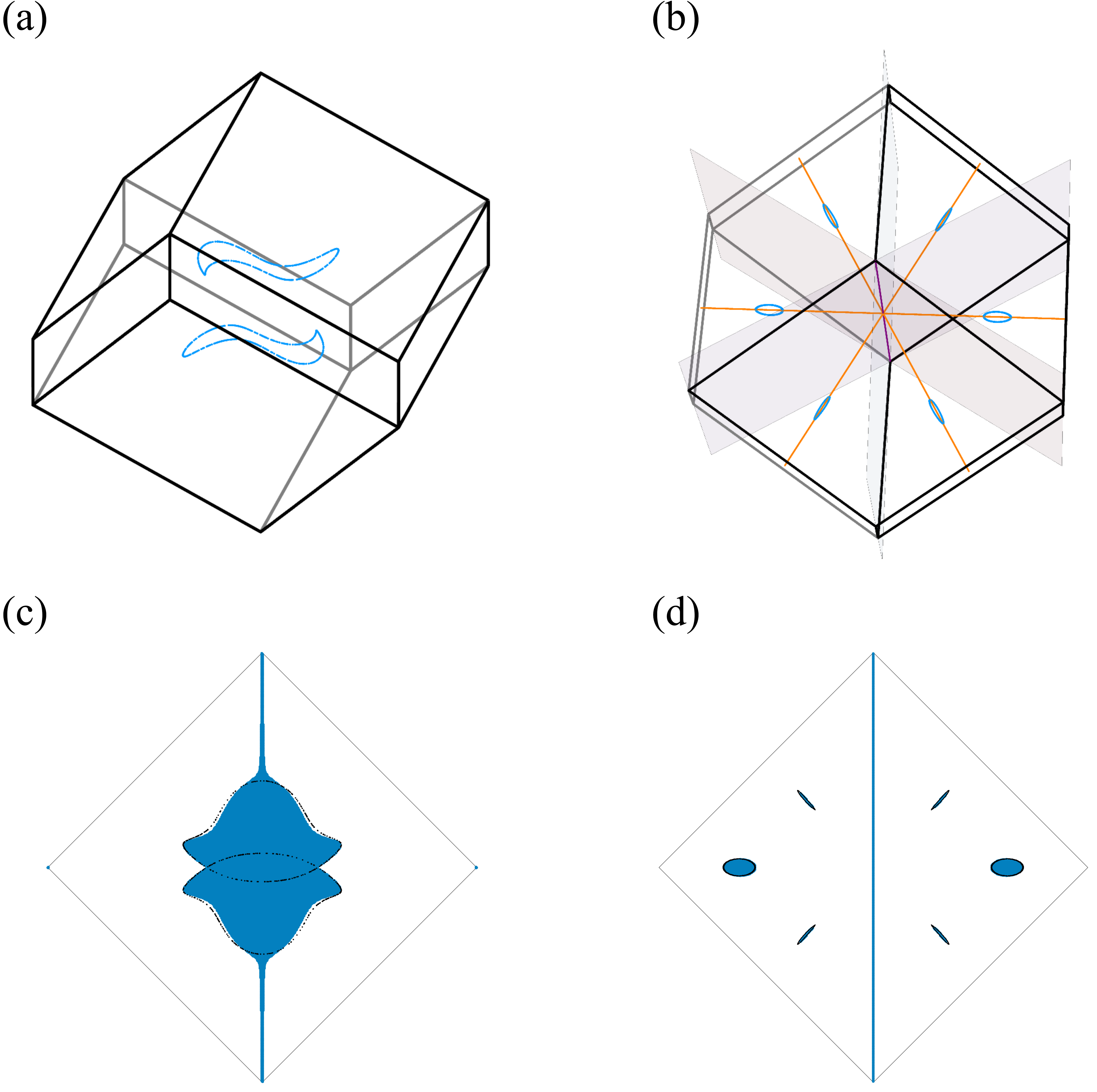}
	\caption{
		Gapless AFII phase. (a) Nodal lines for $(J_x, J_y, J_z) = (0.34, 0.34, 0.32)$ shown in blue. (b) Nodal lines for $(J_x, J_y, J_z) =
		(0.3035, 0.3035, 0.393)$ shown in blue. The shaded planes denote the three mirror invariant planes, whereas the orange lines denote the
		$\textbf{k}$ points invariant under a combination of mirror and inversion symmetries. (c) and (d) Surface states for $J_z = 0.32$ and
		$J_z = 0.393$, respectively, depicted in blue along with projections of bulk nodal lines depicted in black for the 001-surface
		Brillouin zone.
	}
	\label{fig:AFIIPanel1}
\end{figure}

In the AFII phase, as for the SI and AF phases, the presence of a trivially-implemented inversion symmetry prohibits the formation of
stable Fermi surfaces. Furthermore, like for the SI phase, the absence of time-reversal symmetry prevents the formation of nodal lines
protected by sublattice symmetry.

However, it has been shown that in a system where the combination of inversion and time-reversal symmetries is preserved, one may define
both 1D and 2D $\mathbb{Z}_2$ winding numbers~\cite{KimPRL2015,FangPRB2015}. The 1D winding number corresponds to a Berry phase of either 0
or $\pi$ acquired upon traversal of a 1D loop. It may stabilize nodal lines similarly to what has been seen for time-reversal invariant
Kitaev spin liquids, e.g., on the hyperhoneycomb lattice~\cite{obrien_prb_93_2016}. Additionally, pairs of nodal lines can be stabilized by
the presence of a 2D $\mathbb{Z}_2$ winding number, i.e., a nodal line may carry a $\mathbb{Z}_2$ monopole charge~\cite{FangPRB2015}. Due
to their monopole charges, such nodal lines must always be created and annihilated in pairs rather than being continuously deformed to a
point and gapped out in isolation. Crucially, inversion and time-reversal symmetries need not individually be preserved in the system, only
the combination of the two. For the hypernonagon lattice, any fixed flux sector breaks time-reversal symmetry spontaneously. However, in
the AFII phase where one of the inversion symmetries is broken, the combination of the corresponding inversion operation with time-reversal
indeed yields a symmetry of the Hamiltonian. In this case,  line nodes can be stabilized even in the absence of sublattice symmetry.

Diagonalizing the concrete Hamiltonian in the AFII phase reveals an extended gapless region with nodal lines. Restricting the exchange
couplings to the line $J_x = J_y = (1 - J_z)/2$, the gapless portion of the AFII phase runs from $J_z \approx 0.3$, where the ground state
switches from the SI phase to the AFII phase, to $J_z \approx 0.42$, where the ground state switches from the AFII phase to the AF phase,
with a region in between $1/3 < J_z \lesssim 0.392$ for which the spectrum is fully gapped.

At the isotropic point $J_z = 1/3$ there is a fourfold degeneracy at zero energy at the $\Gamma$ point which gaps out as soon as $J_z$ is
increased. However, for $J_z < 1/3$, this fourfold degeneracy is split into two nodal lines which grow larger and move away from each other
as $J_z$ is increased further [see Fig.~\ref{fig:AFIIPanel1}(a)]. Artificially extending the analysis of the AFII flux configuration for
$J_z \lesssim 0.3$, where there is a phase transition to the SI phase, the nodal lines can be seen to wrap around the Brillouin zone before
meeting once more and mutually annihilating at $J_z \approx 0.22$. Putting the system on a slab geometry, one finds both ``drumhead"
surface states~\cite{BurkovPRB2011,PhillipsPRB2014,ChiuPRB2014,ChenNatComm2015,MullenPRL2015,Wawrzik2018} filling the projection of the
nodal lines to the surface Brillouin zone as well as Fermi arc surface states connecting the two projections [see
Fig.~\ref{fig:AFIIPanel1}(c)].  The latter suggests that the nodal lines carry a $\mathbb{Z}_2$ monopole charge~\cite{GorbarPRB2015a,
GorbarPRB2015b}.

For the gapless region corresponding to $J_z \gtrsim 0.392$, a number of nodal lines are stabilized by a 1D $\mathbb{Z}_2$ winding number.
For $J_z \approx 0.392$, a total of six nodal points appear along high-symmetry lines related to one another by inversion and mirror
symmetries or, equivalently, by $C_3$ and inversion symmetries. The high-symmetry lines themselves correspond to momenta invariant under
the combination of mirror and inversion symmetries -- one such line for each of the three mirror planes [see
Fig.~\ref{fig:AFIIPanel1}(b)]. As $J_z$ is increased, the point nodes immediately expand to nodal lines which move through the Brillouin
zone and are heavily deformed. Figure~\ref{fig:AFIIPanel1}(b) shows these nodal lines for $J_z = 0.393$. Due to the severe deformation of
the nodal lines, the figure shows a ``top" view from which the deformation is significantly less evident.

Putting the model on a slab geometry [see Figs.~\ref{fig:AFIIPanel1}(c) and \ref{fig:AFIIPanel1}(d)], there are seen to be drumhead-type
surface states filling the projection of the nodal lines to the surface Brillouin zone as well as the remnant of a Fermi arc from the
$\mathbb{Z}_2$ monopole nodal lines which appear in another part of the phase diagram as discussed above. However, no Fermi arc like states
are observed connecting the disjoint projections of the bulk nodal lines. The lack of such Fermi arc surface states along with the fact
that the bulk nodal lines are not created and destroyed in pairs, rather they appear individually at arbitrary points in the Brillouin
zone, indicates that they carry a 1D winding number rather than the 2D monopole charge.

%
\section{Concluding remarks}
\label{sec:summary_remarks}
%

To summarize, we have explored the physics arising from the crystalline ordering of the $\mathbb{Z}_2$ fluxes in the hypernonagon Kitaev
model. The three-dimensionality of the lattice in combination with existence of odd-length plaquettes gives a potential for at least two
types of finite-$T$ phase transitions -- a topological one originating from loop proliferation of the $\mathbb{Z}_2$ flux excitations and
the spontaneous breaking of time-reversal symmetry leading to a CSL. In addition, a previous study has suggested the breaking of additional
point-group symmetries of the lattice by nonuniform ordering of the $\mathbb{Z}_2$ fluxes. Motivated to find such unconventional
finite-$T$ phase transitions beyond the Landau-Ginzburg-Wilson description, we have employed large-scale, sign-free QMC simulations on the
basis of the kernel polynomial method. While our numerics found a single phase transition, at which time-reversal and lattice symmetries
are simultaneously broken, the transition itself turns out to be strongly first order in nature. Further exploring the zero-temperature
physics of the model, we have established the presence of five distinct nonuniform crystalline $\mathbb{Z}_2$ flux orderings, all of which
except for one are not commensurate with the underlying symmetries of the hypernonagon lattice. In addition, we calculated the band
structures of the itinerant Majorana fermions for all the flux orderings. We found that in the three phases the itinerant Majorana fermions
form both gapped and gapless excitation spectra, and this gapped/gapless boundary does not necessarily  coincide with the boundaries
between the different flux phases. Finally, we clarified the topological nature of the itinerant Majorana band structures by identifying
topological features such as Weyl nodes or nodal lines which are accompanied by topologically protected surface states, Fermi arcs and
drumhead states correspondingly.

\begin{acknowledgments}
P.A.M., Y.K., T.A.B., and Y.M. acknowledge funding by Grant-in-Aid for Scientific Research under Grant No. 15K13533, 16H02206, and
18K03447. Y.M. and Y.K. were also supported by JST CREST (JPMJCR18T2). The Cologne group acknowledges partial support from the Deutsche
Forschungsgemeinschaft (DFG, German Research Foundation), Projektnummer 277146847 -- SFB 1238 (projects C02, C03). M.H. acknowledges
partial funding by the Knut and Alice Wallenberg Foundation and the Swedish Research Council. P.A.M. is supported by JSPS through a
research fellowship for young scientists. The numerical simulations were performed on the supercomputing system in ISSP, the University of
Tokyo, on the CHEOPS cluster at the RRZK Cologne, and on the JUWELS cluster at the Forschungszentrum Juelich.
\end{acknowledgments}

\bibliography{hn_v46}

\appendix

%
\section{Lattice Information}
\label{sec:appendix}
%

As mentioned in Sec.~\ref{sec:lattice}, it is possible to define an equivalent, but deformed version of the hypernonagon lattice in terms
of honeycomb layers joined by mid-bond sites [see Fig.~\ref{fig:fig_1}(b)]. The concrete choice of elementary unit cell for this deformed
lattice has twelve sites with positions given by
\begin{equation}
	\begin{matrix*}[l]
		\mathbf{r}_1    = \left(-\frac{2}{ \sqrt{3}},            0,                  0\right), &
		\mathbf{r}_2    = \left(-\frac{7}{4\sqrt{3}},  \frac{1}{4},                  0\right), \\
		\\
		\mathbf{r}_3    = \left(-\frac{7}{4\sqrt{3}},  \frac{1}{4}, \frac{1}{\sqrt{3}}\right), &
		\mathbf{r}_4    = \left(-\frac{ \sqrt{3}}{2},  \frac{1}{2}, \frac{1}{\sqrt{3}}\right), \\
		\\
		\mathbf{r}_5    = \left(-\frac{1}{2\sqrt{3}}, -\frac{1}{2},                  0\right), &
		\mathbf{r}_6    = \left(-\frac{1}{ \sqrt{3}}, -\frac{1}{2},                  0\right), \\
		\\
		\mathbf{r}_7    = \left(-\frac{1}{ \sqrt{3}}, -\frac{1}{2}, \frac{1}{\sqrt{3}}\right), &
		\mathbf{r}_8    = \left(-\frac{ \sqrt{3}}{2}, -\frac{1}{2}, \frac{1}{\sqrt{3}}\right), \\
		\\
		\mathbf{r}_9    = \left(-\frac{1}{2\sqrt{3}},  \frac{1}{2},                  0\right), &
		\mathbf{r}_{10} = \left(-\frac{1}{4\sqrt{3}},  \frac{1}{4},                  0\right), \\
		\\
		\mathbf{r}_{11} = \left(-\frac{1}{4\sqrt{3}},  \frac{1}{4}, \frac{1}{\sqrt{3}}\right), &
		\mathbf{r}_{12} = \left(                   0,            0, \frac{1}{\sqrt{3}}\right).
	\end{matrix*}
\end{equation}
The lattice vectors are chosen to be
\begin{equation}
	\begin{matrix*}[l]
		\mathbf{a}'_1 = \left(-\frac{\sqrt{3}}{2}, \frac{1}{2}, \frac{1}{\sqrt{3}}\right), &
		\mathbf{a}'_2 = \left(0, -1, \frac{1}{\sqrt{3}}\right), \\
		\\
		\mathbf{a}'_3 = \left(\frac{\sqrt{3}}{2}, \frac{1}{2}, \frac{1}{\sqrt{3}}\right)
	\end{matrix*}
\end{equation}
with corresponding reciprocal lattice vectors
\begin{equation}
	\begin{matrix*}[l]
		\mathbf{q}'_1 = \left(-\frac{2\pi}{\sqrt{3}}, \frac{2\pi}{3}, \frac{2\pi}{\sqrt{3}}\right), &
		\mathbf{q}'_2 = \left(0, -\frac{4\pi}{3}, \frac{2\pi}{\sqrt{3}}\right), \\
		\\
		\mathbf{q}'_3 = \left(\frac{2\pi}{\sqrt{3}}, \frac{2\pi}{3}, \frac{2\pi}{\sqrt{3}}\right).
	\end{matrix*}
\end{equation}

\begin{figure*}[t]
	\centering
	\includegraphics[width=\textwidth]{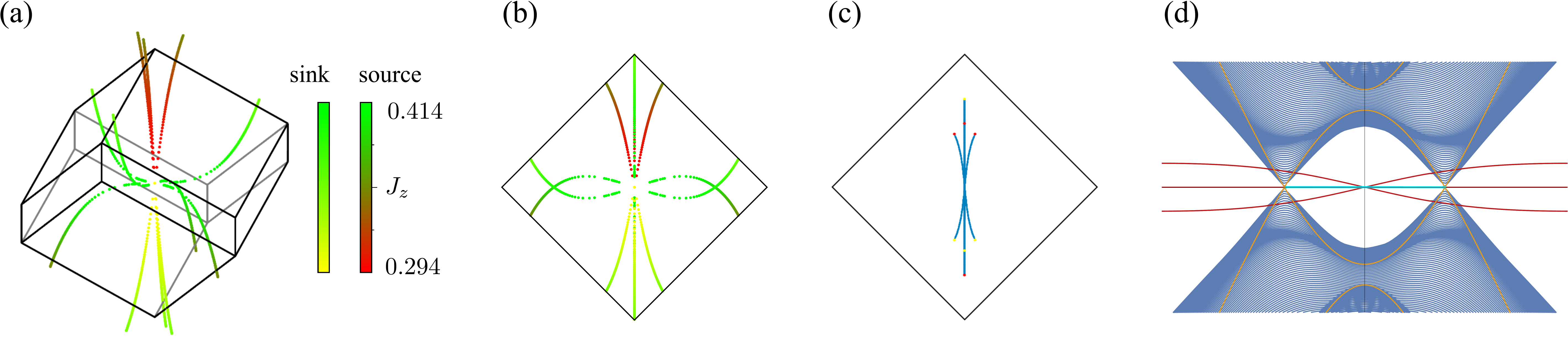}
	\caption{
		Evolution of Weyl nodes for coupling constants $0.294 \leq J_z \leq 0.414$ with $J_x = J_y = (1 - J_z)/2$ in (a) the bulk Brillouin
		zone and (b) projected down to the 001-surface Brillouin zone, respectively. The two Weyl nodes on the $C_3$-invariant axis are not
		shown for clarity. (c) Corresponding Fermi arcs in the 001-surface Brillouin zone for $(J_x, J_y, J_z) = (0.345, 0.345, 0.31)$. This
		includes the Weyl nodes which lie on the $C_3$-invariant axis along with their corresponding Fermi arc. (d) Band structures for
		$J_z = 0.46$. The bulk band structure along the $C_3$-invariant axis is plotted in yellow showing the two Weyl nodes. The band
		structure for the slab geometry along the projection of the $C_3$ axis to the surface Brillouin zone is plotted in blue and red. The
		projection of the two Weyl nodes as well as the Fermi arc which connects them are colored blue.  The bands
		responsible for the remaining surface states left over from the other Weyl nodes, which are gapped out for $J_z \approx 0.414$, are
		colored red.
	}
	\label{fig:AFPanel2}
\end{figure*}

The assignment of $x$, $y$ and $z$ bonds is depicted in Fig.~\ref{fig:fig_1}(b) corresponding to red, green and blue colored bonds,
respectively. Up to permutations of the bond types, this assignment is unique. All $x$ and $y$ bonds are related by a combination of $C_3$
and mirror symmetries. There are two distinct sets of $z$ bonds which are not related by lattice symmetries, however, all bonds of a given
set may be mapped onto each other by a $C_3$ symmetry. The symmetry between $x$ and $y$ bonds is reflected in the ground state phase
diagrams of Fig.~\ref{fig:fig_6}(a).

%
\section{Surface states in the AF phase}
\label{sec:appendix2}
%

As mentioned in the discussion of the AF phase in the main text, in addition to the Fermi arc which terminates at the projection of the
Weyl nodes in the surface Brillouin zone, there appear additional line-like surface states which form incontractible loops. These can be
seen as remnants of Fermi arcs due to Weyl nodes which are gapped out in this part of the phase diagram. If the flux configuration
corresponding to the AF phase is artificially extended beyond its range of validity, i.e., for $J_z \lesssim 0.42$ with $J_x = J_y$, one
finds the existence of an additional six Weyl nodes. These Weyl nodes emerge from the $\Gamma$-point at $J_z \approx 0.294$ and move away
from one another as $J_z$ is increased, always pinned to the three mirror planes. At $J_z \approx 0.414$, the six Weyl nodes once again
meet and annihilate at the $\Gamma$-point. However, as they do so, they trace out incontractible loops across the Brillouin zone. The
evolution (excluding the two Weyl nodes on the $C_3$-invariant axis mentioned in the previous paragraph) in the bulk Brillouin zone as well
as projected down to the 001-surface Brillouin zone, is depicted in Figs.~\ref{fig:AFPanel2}(a) and (b), respectively. 

A consequence of the trajectories of the Weyl nodes through the Brillouin zone is that their corresponding Fermi arcs [pictured in
Fig.~\ref{fig:AFPanel2}(c)], rather than shrinking to a point and gapping out with the Weyl nodes, are stretched out along similar
incontractible loops. The result is that, while the Weyl nodes responsible for the surface states gap out, the surface states themselves
remain behind [as can be seen in Fig.~\ref{fig:AFPanel1}(b) of the main text]. Pictured in Fig.~\ref{fig:AFPanel2}(d) is
a composite of the bulk and slab geometry band structures for $J_z = 0.46$ computed along the $C_3$-invariant axis and its projection to
the 001-surface Brillouin zone, respectively. Here can be seen the Fermi arcs in blue which terminate at the projection of the Weyl nodes
to the surface Brillouin zone at which point they dive back into the bulk. In red, however, are pictured the remnant surface bands from the
Weyl nodes which have been gapped out. These bands are seen to be entirely disconnected from the bulk band structure.

\end{document}